\documentclass[a4paper,12pt]{article}
\usepackage{amsmath,amsfonts,amssymb,cite}
\usepackage[dvips]{graphicx}

\begin{document}

\begin{center}
{\Large\bf The linear radial spectrum of scalar mesons within the QCD sum rules 
in the planar limit\footnote{Presented at QUARKS-2018 (Valday, Russia, May 27 - June 2, 2018).}}
\end{center}
\bigskip
\begin{center}
{S. S. Afonin
and T. D. Solomko}
\end{center}

\begin{center}
{\it Saint Petersburg State University, 7/9 Universitetskaya nab.,
St.Petersburg, 199034, Russia}
\end{center}

\abstract{%
We discuss the radial spectrum of light scalar mesons in the framework of spectral sum rules
in the large-$N_c$ (planar) limit of QCD.
Two methods based on the use of linear radial Regge trajectories are presented.
A special emphasis is placed on the appearance of scalar isoscalar state near 0.5~GeV.
Within the considered sum rules, the existence of this meson
(which could be a large-$N_c$ counterpart to the sigma-meson) is closely
related with existence of resonances in the vector channels and on the radial scalar trajectory.
}
%
%
\section{Introduction}

The physical characteristics of hadrons are encoded in various
correlation functions of corresponding quark currents. One of
the most important characteristics is the hadron mass. The
calculation of a hadron mass from first principles consists in
finding the relevant pole of two-point correlator $\left\langle
JJ\right\rangle$, where the current $J$ is built from the quark
and gluon fields and interpolates the given hadron.
In the real QCD, the straightforward
calculations of correlators are possible only in the framework of
lattice simulations which are still rather restricted.


It is usually believed that confinement in QCD leads to approximately linear
radial Regge trajectories (see, e.g.,~\cite{phen}). The most important
quantity in this picture is the slope of trajectories. The slope is expected
to be nearly universal as arising from flavor-independent non-perturbative
gluodynamics which thereby sets a mass scale for the light hadrons.

Among the phenomenological approaches to the hadron spectroscopy, the method
of spectral sum rules~\cite{svz} is likely the most related with QCD.
In many cases, it permits to calculate reliably the masses of ground
states on the radial trajectories. This method exploits some information from QCD via
the Operator Product Expansion (OPE) of correlation
functions~\cite{svz}. On the other hand, one assumes a certain
spectral representation for a correlator in question. Typically
the representation is given by the ansatz "one infinitely narrow
resonance + perturbative continuum". Such an approximation is very
rough but works well phenomenologically in many
cases. Theoretically the zero-width approximation arises in the large-$N_c$
limit of QCD~\cite{hoof}. In this limit, the
only singularities of the two-point correlation function of 
quark currents $J$ are one-hadron states. For instance,
the two-point correlator of the scalar currents $J=\bar{q}q$ has the following form to
lowest order in $1/N_c$ (in the momentum space),
\begin{equation}
\label{20}
\Pi_S(q^2)=\left\langle J^S(q)J^S(-q)\right\rangle=\sum_n\frac{G_n^2M_S^2(n)}{q^2-M_S^2(n)},
\end{equation}
where the residues appear from the definition of the matrix element $\langle0|J^S|n\rangle=G_nM_S(n)$.
The OPE of the correlator~\eqref{20} in the large-$N_c$
limit and to the lowest order in the perturbation theory reads~\cite{rry}
\begin{equation}
\label{27}
\Pi_S(Q^2)=\frac{3Q^2}{16\pi^2}\log{\frac{Q^2}{\mu^2}}+ \frac{3}{2Q^2}m_q\langle\bar{q}q\rangle
-\frac{\alpha_s}{16\pi}\frac{\langle G^2\rangle}{Q^2}
-\frac{11}{3}\pi\alpha_s\frac{\langle\bar{q}q\rangle^2}{Q^4}+\dots,
\end{equation}
where $\langle G^2\rangle$ and $\langle\bar{q}q\rangle$ denote the
gluon and quark vacuum condensate, respectively. According to the
main assumption of classical QCD sum rules~\cite{svz}, these vacuum
characteristics are universal, i.e., their values do not depend on
the quantum numbers of a quark current $J$ (the method is not
applicable otherwise).

In the present talk we will demonstrate how all these ideas can be used
for calculation of large-$N_c$ masses of light scalar mesons.

\section{Scalar sum rules: Some results}

We will assume the linear radial spectrum with universal slope
\begin{equation}
\label{21}
M_S^2(n)=\Lambda^2(n+m_s^2), \qquad n=0,1,2,\dots,
\end{equation}
and (for consistency with the OPE): $G_n=G$.
With the linear ansatz~\eqref{21} for
the radial mass spectrum, the expression~\eqref{20} can be summed analytically,
expanded at large $Q^2=-q^2$ and compared with the corresponding
OPE in QCD. Thus one obtains a set of sum rules. Similar large-$N_c$ sum rules were considered many
times in the past for vector, axial, scalar and pseudoscalar channels (see, e.g., Refs. in~\cite{sr}).

As {\it apriori} we do not know reliably the radial Regge behavior
of scalar masses, two simple possibilities can be considered: (I)
The ground $n=0$ state lies on the linear trajectory~\eqref{21};
(II) The state $n=0$, below called $\sigma$, is not described by
the linear spectrum~\eqref{21}. The second assumption looks more
physical. Within the latter assumption, the mass of $\sigma$-meson
can be derived as a function of the intercept parameter $m_s^2$
(we refer to Ref.~\cite{we1} for details, the chiral limit is considered),
\begin{equation}
\label{33}
M_\sigma^2=\frac{\frac{1}{16\pi^2}\Lambda^6m_s^2\left(m_s^2+\frac12\right)\left(m_s^2+1\right)+
\frac{11}{3}\pi\alpha_s\langle\bar{q}q\rangle^2}
{\frac{3}{32\pi^2}\Lambda^4\left(m_s^4+m_s^2+\frac16\right)+\frac{\alpha_s}{16\pi}\langle G^2\rangle}.
\end{equation}
Substituting the physical values of vacuum condensates and numerical value for slope $\Lambda^2$ obtained
from a solution of QCD sum rules, the mass function~\eqref{33} is displayed in Fig.~1~\cite{we1}.
\begin{figure}[ht]
\center{\includegraphics[width=0.7\linewidth]{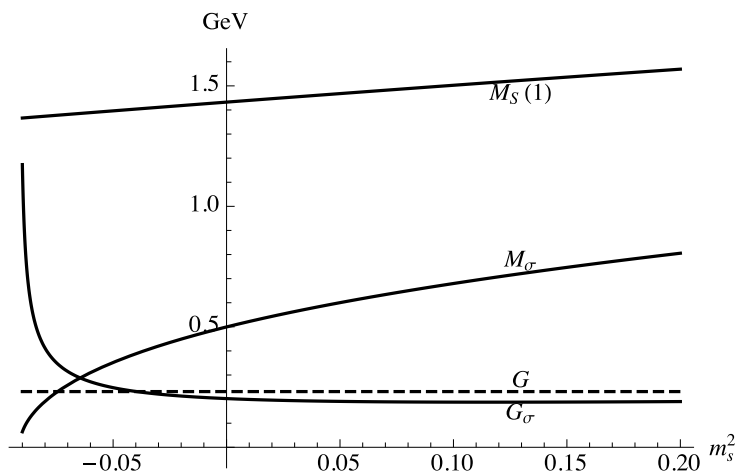}}
\vspace{-0.3cm}
\caption{\small The values of $M_\sigma$, $G_\sigma$, $G$, and
the first state on the scalar trajectory $M_S(1)$ as a
function of dimensionless intercept $m_s^2$.}
\end{figure}

The mass of the first radially excited state $M_S(1)$ is rather stable and
seems to reproduce the mass of $a_0(1450)$-meson,
$M_{a_0(1450)}=1474\pm19$~MeV~\cite{pdg}. Its isosinglet partner
(the candidates is $f_0(1370)$) should be degenerate with
$a_0(1450)$ in the planar limit.

The plot in Fig.~1 demonstrates that the actual prediction for
$M_\sigma$ is rather sensitive to the intercept of scalar linear
trajectory, though initially $M_\sigma$ is not described by the
linear spectrum~\eqref{21}. And {\it vice versa}, the expected
value of $M_\sigma$ (around $0.5$~GeV~\cite{pdg}) imposes a strong
bound on the allowed values of intercept $m_s^2$. The plot in
Fig.~1 shows that $m_s^2$ is likely close to zero.

Thus, interpolating the scalar states by the simplest quark bilinear current,
we predict (in the large-$N_c$ limit!) a light scalar resonance with mass about $500\pm100$~MeV
which could be a reasonable candidate for the scalar sigma-meson $f_0(500)$~\cite{pdg}.

\section{Borelized scalar sum rules: Some results}

The original QCD sum rules made use of the Borel transformation~\cite{svz},
\begin{equation}
\label{6}
L_M\Pi(Q^2)=\lim_{\substack{Q^2,n\rightarrow\infty\\Q^2/n=M^2}}\frac{1}{(n-1)!}(Q^2)^n\left(-\frac{d}{dQ^2}\right)^n\Pi(Q^2),
\end{equation}
The borelized version has a number of advantages and can be applied to
our large-$N_c$ case. The details are contained in Ref.~\cite{we2}.
In short, the mass of ground scalar meson $m_0\equiv M_S(0)$ as a function of Borel parameter
is shown in Fig.~2. It is seen that there are two solutions with "Borel window" extending to infinity.
\begin{figure}[ht]
    \center{\includegraphics[scale=0.7]{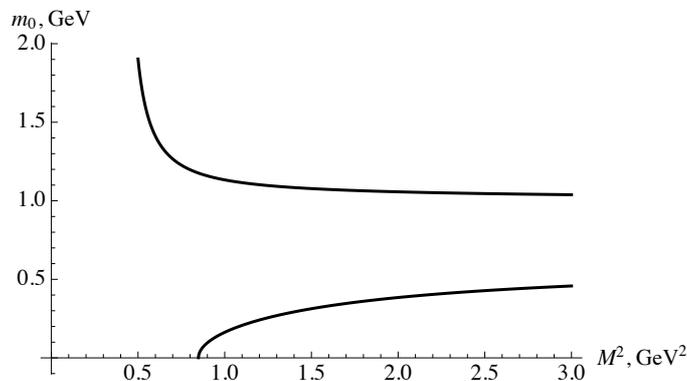}}
    \vspace{-0.3cm}
    \caption{\small The mass of ground scalar meson $m_0\equiv M_S(0)$ as a function of Borel parameter at $\Lambda^2=1.38\,\text{GeV}^2$~\cite{we2}.}
\end{figure}
The corresponding asymptotic values are given by
\begin{equation}
\label{11}
M_S^2(0)=\frac{\Lambda^2}{2}\pm\frac{1}{2}\sqrt{\frac{\Lambda^4}{3}-64\pi^2\left(m_q\langle\bar{q}q\rangle + \frac{\alpha_s}{24\pi}\langle G^2\rangle\right)}.
\end{equation}

The heavier state corresponds to the ground scalar mass in the standard QCD sum rules.
Normalizing this mass to the value $m_{f_0}=1.00\pm0.03$~GeV extracted from these canonical sum rules~\cite{rry}
we predict the value of slope for the scalar trajectory, $\Lambda^2_{f_0}=1.38\pm0.07$~GeV$^2$, which is
used in Fig.~2. We obtain then the mass of the lightest scalar state, $M_\sigma\approx0.62$~GeV.

We arrive thus at the conclusion that our method predicts two
parallel scalar trajectories. The ground state on the first trajectory
can be identified with $f_0(980)$ and on the second one with $f_0(500)$~\cite{pdg}.
The existence of two parallel radial scalar trajectories seems to agree
with the experimental data~\cite{phen}. The masses of predicted radial
states and a tentative comparison with the observed scalar mesons for two
trajectories are displayed in Tables~1 and~2, correspondingly.

\begin{table}[ht]
\caption{\small The radial spectrum of the first $f_0$-trajectory for the slope $\Lambda^2=1.38\pm0.07\,\text{GeV}^2$.
The first 5 predicted states are tentatively assigned to the resonances $f_0(980)$, $f_0(1500)$,
$f_0(2020)$, $f_0(2200)$, and $X(2540)$~\cite{pdg}.}
\begin{center}
$\begin{array}{|c|c|c|c|c|c|}
 \hline
 n & 0 & 1 & 2 & 3 & 4\\
 \hline
 m_{f_0}\,\text{(th 1)} & 1000\pm30 & 1540\pm20 & 1940\pm40 & 2270\pm50 & 2560\pm50 \\
 \hline
 m_{f_0}\,\text{(exp 1)} & 990 \pm 20 & 1504 \pm 6 & 1992 \pm 16 & 2189 \pm 13 & 2539 \pm 14^{+38}_{-14} \\
 \hline
\end{array}$
\end{center}
\end{table}

\begin{table}[ht]
\caption{\small The radial spectrum of the second $f_0$-trajectory for the slope $\Lambda^2=1.38\pm0.07\,\text{GeV}^2$.
The first 5 predicted states are tentatively assigned to the resonances $f_0(500)$, $f_0(1370)$,
$f_0(1710)$, $f_0(2100)$, and $f_0(2330)$~\cite{pdg}.}
\begin{center}
$\begin{array}{|c|c|c|c|c|c|}
 \hline
 n & 0 & 1 & 2 & 3 & 4\\
 \hline
 m_{f_0}\,\text{(th 2)} & 620 & 1330\pm30 & 1780\pm40 & 2130\pm50 & 2430\pm60 \\
 \hline
 m_{f_0}\,\text{(exp 2)} & 400\text{--}550 & 1200\text{--}1500 & 1723^{+6}_{-5} & 2101 \pm 7 & 2300\text{--}2350 \\
 \hline
\end{array}$
\end{center}
\end{table}

\section{Discussions and conclusions}

We have put forward new extensions of SVZ sum rules making use of the large-$N_c$ (planar) limit
and assuming for the radial excitations a linear Regge spectrum with
universal slope. The choice of spectrum is motivated by hadron string
models and related approaches and also by the meson spectroscopy.
The considered ansatz allows to solve the arising sum rules with a
minimal number of inputs.

The prediction of the second scalar trajectory is a rather
surprising feature of borelized planar sum rules~\cite{we2}. The ground
state on the second radial trajectory turns out to be significantly
lighter than on the first trajectory. It looks tempting to identify
this state with the elusive $\sigma$ (called also $f_0(500)$) meson~\cite{pdg}.
The lightest scalar state in the standard SVZ sum rules lies near 1~GeV~\cite{rry}
and cannot be made significantly lighter within this method~\cite{sigma}.
Our extension of the SVZ method leads thus to a new result. It is interesting to
check whether a similar result appears in the framework of unborelized
planar sum rules. Our analysis in the first part gives a positive answer.
A light scalar state near 0.5~GeV, however, emerges in a
different way~\cite{we1}.

When one predicts some quark-antiquark state it is important to
indicate its place on the angular Regge trajectory as well. In
other words, what are $f_2,\,f_4,\,\dots$ companions of $f_0(500)$
on this trajectory? In order to answer this question we must know
the slope of the trajectory under consideration. According to the
analysis of the first paper in Ref.~\cite{phen}, the slope of $f_0$ trajectory, most
likely, lies in the interval $1.1\div1.2$ GeV$^2$. Several
independent estimations made in some further papers of
Ref.~\cite{phen} seem to confirm this value. Consider for example the estimate
on the $\sigma$-meson mass,
$m_\sigma\approx390$~MeV~\cite{we1}. Then we obtain
$m_{f_2}\approx1.53\div1.60$~GeV. The PDG contains a well
established resonance $f_2(1565)$~\cite{pdg} with mass
$m_{f_2(1565)}=1562\pm13$~MeV. It is a natural companion of
$\sigma$-meson on the corresponding angular Regge trajectory. The
next state would have the mass $m_{f_4}\approx2.13\div2.23$~GeV.
The discovery of the predicted tensor meson $f_4$ (and perhaps the
next companion $f_6$ with $m_{f_6}\approx2.60\div2.71$~GeV) would
confirm our conjecture about the form of Regge trajectory with the
$\sigma$-meson on the top. A tentative candidate for our $f_4$ in
the Particle Data is the resonance $f_J(2220)$ having still
undetermined spin --- its value is either $J=2$ or
$J=4$~\cite{pdg}. Our model would favor the second possibility.

It is interesting to note that the predicted trajectory is drawn
in the first paper of Ref.~\cite{phen} among numerous angular Regge trajectories for
isosinglet $P$-wave states of even spin. But the resonance
$f_2(1565)$ is replaced there by $f_2(1525)$ (and is absent on
other trajectories). As a result, $m_{f_0}^2$ has a very small
negative value leading to disappearance of a scalar state from
this trajectory. The predicted $f_4$-companion is labelled as
$f_4(2150)$. The modern PDG contains the state
$f_2'(1525)$ but this resonance is typically produced in reactions
with $K$-mesons that evidently indicates on the dominant strange
component. For this reason we should exclude it from our
estimates.

Our prediction of the Regge trajectory containing the
$\sigma$-meson on the top seems to contradict to studies of the
$\sigma$-state on the complex Regge trajectory which claim that
because of very large width the corresponding state cannot belong
to usual Regge trajectories~\cite{pelaez}. It is not
excluded, however, that this observation may simply indicate on
limitations of the usual methods which are applied to description
of the $\pi\pi$-scattering. These methods are based on analyticity
and unitarity of $S$-matrix and do not contain serious dynamical
inputs. The generation of a huge width for $f_0(500)$ represents,
most likely, some dynamical effect. For this reason uncovering the
genuine nature of $\sigma$-meson requires dynamical approaches.
Such approaches should simultaneously describe all resonances on the
radial scalar trajectory and resonances in
other channels (vector, tensor {\it etc.}). The planar QCD sum rules
effectively take this into account and the appearance of $\sigma$
is directly related with the existence of all these resonances. The
given property constitutes a conceptual difference of considered
method from the dispersive approaches.

Since the used sum rule method is based on the narrow-width
approximation, a direct translation of our predictions to the
physical parameters of a broad resonance may look questionable. As a
matter of fact, we claim only that a scalar isoscalar pole in the
range 400--600 MeV can naturally exist in the large-$N_c$ limit.

Another pertinent question is why the $\sigma$-meson lies below
the linear radial Regge trajectory like the ground vector states?
In the latter case, one can propose a simple qualitative
explanation. The ground vector states are $S$-wave, so they
represent relatively compact hadrons. In this case, a contribution
from the Coulomb part of the Cornell confinement potential,
$V(r)=-\frac43\frac{\alpha_s}{r} + \sigma r$, is not small,
effectively "decreasing" the tension $\sigma$ at smaller distances
and, hence, masses of the ground $S$-wave states. In the case of
$\sigma$-meson, one can imagine the following situation: This
state represents a tetraquark but the admixture of additional
$q\bar{q}$-pair is relatively small and gives a non-dominant contribution to
the mass. For this reason we may use the large-$N_c$ limit as the
first approximation. However, due to the extra $q\bar{q}$-pair,
the $\sigma$-meson (originally representing a scalar $P$-wave
state) can exist as a $S$-wave state. Due to this phenomenon, on
the one hand, the decay of this state becomes OZI-superallowed,
explaining thereby its abnormally large width, on the other hand,
its mass decreases similarly to the masses of ground $S$-wave
vector mesons.

In conclusion, our analysis demonstrates that the existence of a light
scalar state is well compatible with the structure of the planar
sum rules in the scalar channel and may follow in a natural way
from the Regge phenomenology in the large-$N_c$ limit.

\end{document}